\documentclass[11pt]{article}
\usepackage{graphicx}
\usepackage[margin=1.15in]{geometry}
\usepackage[usenames,dvipsnames]{color}
\usepackage{url}
\usepackage[colorlinks = true,
            linkcolor = blue,
            urlcolor  = blue,
            citecolor = blue,
            anchorcolor = blue]{hyperref}

\newcommand\pubnumber{Transcendental Preprint }
\newcommand\pubdate{\today}

\def\Title#1{\begin{center} {\LARGE #1 } \end{center}}
\def\Author#1{\begin{center}{ \sc #1} \end{center}}
\def\Address#1{\begin{center}{ \it #1} \end{center}}

\newcommand\pubblock{\rightline{\begin{tabular}{l} \pubnumber\\
         \pubdate \end{tabular}}}
\newenvironment{Abstract}{\begin{quotation} \begin{center}
                       ABSTRACT
     \end{center}\bigskip  }{\end{quotation}}

\def\beq{\begin{equation}}
\def\eeq#1{\label{#1}\end{equation}}
\def\eeqn{\end{equation}}

\newenvironment{Eqnarray}%
   {\arraycolsep 0.14em\begin{eqnarray}}{\end{eqnarray}}
\def\beqa{\begin{Eqnarray}}
\def\eeqa#1{\label{#1}\end{Eqnarray}}
\def\eeqan{\end{Eqnarray}}

\let\bar=\overbar

\def\lsim{\mathrel{\raise.3ex\hbox{$<$\kern-.75em\lower1ex\hbox{$\sim$}}}}
\def\gsim{\mathrel{\raise.3ex\hbox{$>$\kern-.75em\lower1ex\hbox{$\sim$}}}}

\def\del{\partial}
\def\Dslash{\not{\hbox{\kern-4pt $D$}}}
\def\dslash{\not{\hbox{\kern-2pt $\del$}}}
\def\pslash{\not{\hbox{\kern-2pt $p$}}}
\def\ETmiss{\not{\hbox{\kern-4pt $E$}}_T}

\def\Dlr{\mathrel{\raise1.5ex\hbox{$\leftrightarrow$\kern-1em\lower1.5ex\hbox{$D$}}}}

\def\MSB{{\bar{M \kern -2pt S}}}
\def\msb{{\bar{\scriptsize M \kern -1pt S}}}

\def\drb{{\bar{\scriptsize D \kern -1pt R}}}

\newcommand\snowmass{\begin{center}\rule[-0.2in]{\hsize}{0.01in}\\\rule{\hsize}{0.01in}\\
\vskip 0.1in Submitted to the  Proceedings of the US Community Study\\ 
on the Future of Particle Physics (Snowmass 2021)\\ 
\rule{\hsize}{0.01in}\\\rule[+0.2in]{\hsize}{0.01in} \end{center}}

\begin{document}

\pubblock

\Title{Analysis Description Language: A DSL for HEP Analysis}

\bigskip 

\Author{Harrison B. Prosper\textsuperscript{1}, Sezen Sekmen\textsuperscript{2}, Gokhan Unel\textsuperscript{3}}

\medskip

\Address{\textsuperscript{1}Florida State University, USA \\
\textsuperscript{2}Kyungpook National University, Korea \\
\textsuperscript{3}University of California, Irvine, CA, USA}

\medskip

 \begin{Abstract}
\noindent 

We propose to adopt a declarative domain specific language for describing the physics algorithm of a high energy physics (HEP) analysis in a standard and unambiguous way decoupled from analysis software frameworks, and argue that this approach provides an accessible and sustainable environment for analysis design, use and preservation. Prototype of such a language called Analysis Description Language (ADL) and its associated tools are being developed and applied in various HEP physics studies. We present the motivations for using a DSL, design principles of ADL and its runtime interpreter CutLang, along with current physics studies based on this approach. We also outline ideas and prospects for the future. 
Recent physics studies, hands-on workshops and surveys indicate that ADL is a feasible and effective approach with many advantages and benefits, and offers a direction to which the HEP field should give serious consideration.

\end{Abstract}

\snowmass

\begin{center}{
COMPF5: End user analysis \\
COMPF7: Reinterpretation and long-term preservation of data and code}
\end{center}

\def\thefootnote{\fnsymbol{footnote}}
\setcounter{footnote}{0}

\newpage

\section{The framework dilemma}

The High Energy Physics (HEP) community is engaged in a global effort to discover new physics to take us beyond the standard model. To date, however, no compelling experimental evidence exists indicating which, if any, of the current theoretical ideas are the best ones to pursue. In the absence of firm guidance, physicists at the CERN Large Hadron Collider (LHC) and other HEP experiments recognize the necessity of analyzing HEP data in as many ways as possible in the hope that no stone has been left unturned. 
There is, moreover, a constant flow of ideas between the experimental and phenomenology communities on how to probe new models and unexplored final states. A central requirement for these ambitious collaborative efforts to be successful is not only the effectiveness of the data analyses themselves but also the effectiveness of the communication of the ideas they embody.

The collaborative efforts can be made more effective by building analysis environments that are highly accessible, easy to use, and in which ideas are readily communicated to others.  This will lead to broader participation in HEP analysis and increase the diversity of ideas and studies. In this White Paper, we discuss an innovative approach to analysis environments that we argue should be vigorously pursued by HEP physicists. 

Analyses in HEP require processing independent ensembles of data products called {\it events} from real and simulated particle collisions. The physics algorithm in this processing includes defining analysis objects, defining quantities based on event properties, selecting events, re-weighting simulated events to improve their agreement with real collision events, estimating backgrounds and interpreting experimental results by comparing them to predictions. 

These tasks are usually performed within analysis frameworks.  The use of frameworks is the norm because of the many resources they provide to the analyst including access to large data sets, calibration constants and large-scale computing. Analysis frameworks are written in general purpose languages (GPL), such as C++ and Python, in order to benefit from the flexibility and power of these languages and their extensive user support infrastructure.  However, using frameworks can erect technical and conceptual hurdles. The technical hurdle is the need to acquire software and system level expertise. The more serious hurdle is the tendency for the physics algorithms constituting an analysis to be scattered throughout the framework and obscured by technical complexities. The descriptions of physics algorithms are generally written in the same GPL as the frameworks. However, the flexibility and power that make a GPL ideal for writing frameworks invariably yield physics algorithm descriptions that are difficult to decipher, maintain, communicate, and reason about.

Moreover, the physicist is often faced with the multiple frameworks problem, the need to work with different research groups who invariably use frameworks that differ in their details even when identical concepts are being expressed. While there is a move towards fewer and better maintained frameworks, we argue that a further step can, and should, be taken to allow a physicist to express ideas more directly and in ways that are more natural.  We argue that the key step to be taken is decoupling the description of the analysis from the software framework in which it runs and presenting the analysis in a standardized way.


\section{A solution proposal}

The dilemma arising from the need to use analysis frameworks and the need to reduce the complexity experienced by the analyst can be resolved by using a  \emph{domain specific language} (DSL) for describing the physics content of analyses. The DSL would be customized to express analysis-specific concepts and reflect the conceptual reasoning of particle physicists.
The utility of a DSL would be further enhanced by taking a declarative approach, i.e. having the DSL express only the logic of a computation without describing its control flow. This paradigm-shifting idea has started to gain interest in the HEP community in the recent years, and several DSL approaches are studied, which led to various working prototypes~\cite{Sekmen:2020vph}. 

We should note however that while the focused discussions and developments of DSLs for analysis are relatively new, the HEP community has, for many years, already been using DSLs embedded within the ROOT framework\,\cite{BRUN199781} under the guise of {\tt TTreeFormula}, {\tt TTree::Draw} and {\tt TTree::Scan}, which allow visual or textual representation of {\tt TTree} contents for simple and quick exploratory analysis. Though these  include only arithmetic operations, mathematical functions and basic selection criteria, they have served as highly practical tools for analysis.  

There are two main approaches in designing DSLs.  Embedded DSLs rely on the syntax of an existing GPL, while external DSLs have a custom syntax of their own, more tailored to the semantics of the context they describe.  For analysis in HEP, both approaches have been considered.

For embedded DSLs, one early example is the {\tt F.A.S.T.} framework\,\cite{FAST} that incorporates a YAML-based declarative DSL for analysis description that manages both analysis definitions and data processing. It is used in CMS, LZ and DUNE analyses. The more recent DSL developments are based on Python. For example {\tt NAIL} (Natural Analysis Implementation Language)~\cite{NAIL} is a declarative Python-based embedded DSL with a syntax resembling that of ROOT framework's  RDataFrame (RDF)~\cite{Piparo:2699587}, and that can be converted to an RDF-based C++ code. It has been used in the CMS analyses that observed the first evidence for Higgs to two muon decays~\cite{CMS:2020xwi}. Another Python-based DSL linked to RDF is defined within the {\tt bamboo} framework~\cite{David:2021ohq}, which is used in CMS for several analyses including Phase-2 studies for Snowmass.  One more declarative example within Python is {\tt FuncADL}~\cite{Proffitt:2021wfh}, which explores functional programming and query languages such as {\tt LINQ}~\cite{linq}, and is inspired by a former infrastructure called {\tt LINQtoROOT}~\cite{Watts:2015saa}.

The embedded DSL approach benefits from the guidance of an already existing syntax, the potential prior familiarity of the user community with the rules of that syntax, and the existence of infrastructures to make that syntax executable. However, using a GPL syntax would not be completely immune to the issue of intermixing the physics algorithms and technical operations. We therefore advocate the external DSL approach as a more robust way of isolating the physics information.

On the external DSL side, the most developed example is ``Analysis Description Language", or shortly {\tt ADL}~\cite{adlweb, Unel:2021edl, Prosper:2021umk}, which is a declarative DSL describing the physics content of a HEP analysis in a standard and unambiguous way in a framework-independent manner.
There also exists a toy language called {\tt PartiQL}, designed to demonstrating features that would be a radical departure from general purpose languages, addressing problems specific to particle physics~\cite{PartiQL}. It is extended in {\tt AwkwardQL}, designed to perform set operations on data expressed as awkward arrays~\cite{awkward}. 

An external DSL describing HEP data analysis must satisfy at least three requirements: 1) it must be natural to a physicist, 2) it must be unambiguous, and 3) it must be domain complete, meaning that it can describe a very large fraction of (and ideally all) conceivable analyses now and throughout the LHC era. To be useful, the language must satisfy a 4th requirement: the possibility to render it executable, preferably in an automated fashion.  

Adapting a DSL, in particular, an external DSL, would yield many advantages. 
First, a DSL makes analysis writing explicit and clear. It can be designed to avoid potential ambiguities -- e.g., when reconstructing a Z boson, whether one selects the opposite-sign, same-flavor lepton pair within the Z mass range with the highest transverse momentum leptons or the pair closest to the Z mass; which isolation criteria are used for which muon definition; which version of an an event variable calculation is used in the control region, etc.
Moreover, a HEP-specific DSL would allow the rapid prototyping of analyses and dramatically reduce the time needed to test ``what-if'' ideas as well analysis sensitivity and optimization studies.  
Framework-independence would also make it straightforward to write, test, compare and document alternative object or event selections and their performances in parallel in an organized way. 
This feature can be further generalized by building tools that automatically generate large sets of optimal analyses to explore the huge variety of final states proposed by current and future models of new physics. 

Second, having a standard DSL using HEP semantics  will make analyses self-documenting and, therefore, easier to communicate their physics content, thus facilitating their validation and review by others. This could include communication within the analysis team, with reviewers and referees, or with others who wish to understand the analysis.  The DSL would also serve to communicate analysis ideas and practices between different communities, such as between different experiments or between experimentalists and theorists.  For example, an experimentalist could easily communicate object definitions to a theorist, while a theorist could develop a discriminating variable that depends on properties of common objects that could be directly used by an experimentalist. 

Third, adapting a DSL would facilitate the (re)interpretation of analysis results by the analysis teams, physicists from other experiments or theorists. In particular, (re)interpretation  studies by phenomenologists require reimplementations of analyses, which can be very difficult if the published information is inadequate as reported by the community~\cite{Abdallah:2020pec, Kraml:2012sg}. Ready access to complete, accurate descriptions of analyses that can be executed in multiple frameworks will enable correct re-execution of published analyses.

Moreover, the ability to describe an analysis in a framework-independent manner simplifies analysis preservation beyond the lifetime of analysis frameworks and of the experiments. The preservation aspect of DSLs is also discussed in a dedicated Snowmass White Paper contribution on data and analysis preservation~\cite{snowmass:preservation}.
A preserved repository of analyses written in a standard, easy-to-learn DSL would also constitute a \emph{learning database} for students or other physics enthusiasts. Such a resource would permit learning by example from a wide spectrum of physics analyses that may inspire new analysis ideas.  


A framework-independent description of a physics analysis with an external DSL would also facilitate multi-purpose use of the description. The description could be automatically translated or incorporated into the framework most suitable for a given purpose. For example, runtime interpretation may be suitable for rapid analysis prototyping or for training students; automatic translation to C++ could be relevant for memory efficiency; translation to Python could capitalize on its extensive scientific computing ecosystem; or translation to YAML may be more relevant for analysis queries. A framework-independent description, underpinned by advanced compiler technology, would make it possible to map a DSL into future GPLs and frameworks. 

In the following, we will focus on ADL, which was designed and implemented by our team, and provide more details on the design principles and physics applications along with ideas and prospects for the future.

\section{ADL and CutLang - The Present}

ADL originated from a series of discussions at the Les Houches Physics at TeV Colliders Workshop in 2015, where a group of experimentalists and theorists (including the authors) agreed that a standardized way to describe LHC analyses could be of great benefit to the community.  Motivations, use cases, and features that would inform the design of such a language were debated and resulted in the first prototype of an external DSL called the ``Les Houches Analysis Description Accord ({\tt LHADA})"\,\cite{Brooijmans:2016vro}. At the same time, a parallel effort called {\tt CutLang}~\cite{Sekmen:2018ehb, Unel:2021edl, Unel:2019reo} was exploring a runtime interpretable external DSL that provides an easy analysis environment, in particular for beginner students.  The LHADA and CutLang languages were based on the same design principles, and had similar syntax, and thus were merged in 2019, combining the best of both into a single prototype DSL called {\bf ADL}~\cite{adlweb, Unel:2021edl, Prosper:2021umk}. ADL has been designed focused on LHC physics, it should be principally applicable to data analysis in other HEP experiments.  

ADL is a human-readable language written in a plain text file, called an ADL file, that contains the physics algorithm description. The ADL file consists of multiple types of blocks rendering a clear separation between different analysis components, including object, variable, and event selection definitions. Blocks have a keyword-expression structure, where keywords specify analysis concepts and operations.  ADL uses block names and keywords that are meaningful in the HEP domain, such as {\tt object}, {\tt region}, {\tt reject}, {\tt select}, {\tt weight}. 
An \texttt{object} block defines the processing to be applied to a collection of objects. 
A \texttt{region} defines a filter that is applied to the {\tt event}. The output of a \texttt{region} is {\tt true} or {\tt false}, that is, whether the current event is to be kept or discarded. 
Weights other than 0 and 1 can also be assigned to {\tt object}s and {\tt region}s using the keyword {\tt weight}.
ADL includes mathematical and logical operations, comparison and optimization operators, collection reducer operators, 4-vector algebra and standard HEP functions (e.g. $\Delta\phi$, $\Delta R$).  One- and 2-dimensional histograms for object and event quantities can also be defined.  Preexisting selection results including event counts and associated uncertainties (e.g., those published by experiments) can be documented in an ADL file. 

One should note however that some analyses may contain variables which are impractical or impossible to describe directly with the ADL syntax.  These include quantities calculated with complicated algorithms (such as complex kinematic variables) or non-analytical functions (such as machine learning models).  ADL's method is to encapsulate these in self-contained external functions written with a GPL, and reference those functions from within an ADL file.  The functions would be stored in a central, accessible database. Though this method excludes part of the analysis details from the ADL file, it keeps intact the clear and organized structure for the description.

Two approaches have been studied to render ADL executable. One is the transpiler approach, where an analysis written in ADL syntax is converted into a GPL. An example is the prototype {\tt adl2tnm} transpiler, where a Python script translates ADL into C++ code that can be compiled into an executable program~\cite{Brooijmans:2018xbu}. The other approach, explored by {\tt CutLang}, is runtime interpretation, where ADL can be directly executed, without the need of intermediate translation or compilation. CutLang is written in C++ and and is based on ROOT classes. It performs automatic ADL parsing using Lex \& Yacc. The interpreter is extended with a framework to manage input/output operations.  It can automatically recognize and process multiple input event formats commonly used in HEP. CutLang is available in multiple platforms including Docker, Conda and Jupyter.  Technical description and capabilities of CutLang, which is the most complete infrastructure for processing ADL, is provided in full detail in~\cite{Sekmen:2018ehb, Unel:2019reo, Unel:2021edl}. Transpiler and runtime interpreter infrastructures are collectively referred to as \emph{compilers}. 

These compilers are embedded in frameworks that manage the analysis workflow. For example, CutLang may refer to both the compiler/interpreter and the surrounding framework. In a typical ADL workflow, the system takes as input the ADL file, external functions (when required) and events in ROOT format. The output includes cutflows, counts, histograms, and optionally, sets of selected events.

Current versions of ADL and CutLang are already being used to perform physics studies. Tens of published LHC analyses from from different areas (e.g. especially from supersymmetry and exotics but also from top and Higgs physics), including several CMS Phase-2 analyses performed for Snowmass 2021, have been studied and implemented with the ADL syntax.  Some of these analyses, which can be already processed by CutLang, are published in a github database~\cite{adllhcanl}. 
The database can be used as a physics information source, serve reinterpretation studies as well as analysis queries, comparisons and combinations. It is intended as a preliminary step towards long-term analysis preservation. 
ADL and CutLang were also used in a Future Circular Collider (FCC) sensitivity study~\cite{Paul:2020mul} and proved to be a very practical approach that could be employed in similar future projection studies. Moreover, they are also used in schools for training students in HEP analysis~\cite{Adiguzel:2020brl}. Additionally, work is in progress to establish ADL/CutLang as an accessible analysis model for ATLAS and CMS Open Data.  However, the most important physics goal is to achieve the capability of performing a full-fledged LHC data analysis with ADL and tools (in particular, CutLang).  Work has already started in this direction.

\section{ADL and CutLang - The Future}

The prototyping of ADL and CutLang along with the physics studies described above demonstrated the feasibility, effectiveness and potential of using an external DSL for HEP data analysis. 
The field tests also indicated that analysts with different levels of skill find ADL and CutLang easy to learn and use. 
However there is still much to accomplish until we arrive at a fully domain complete language capable of expressing more than 90\% of known analyses (including those based on machine learning).  Similarly, infrastructures such as CutLang need to be adapted to incorporate all language extensions and to comply with all requirements of large scale data analysis. We should strongly emphasize that a realistically useful DSL can only be built upon implementing a large variety of analysis examples, and that the current prototypes immensely benefited from this approach.  Therefore physics applications should progress in parallel, and an iterative design approach would be the most effective.  In the following, we present some ideas and prospects for the future of DSL design, derived from our experience with ADL and CutLang.

\subsection{Extending the language scope}

While the existing ADL prototype can already express simple analysis operations such as object and event selections, or basic variable definitions, it needs to be extended. ADL, or any other candidate for a domain-complete DSL would need to include syntax representations for the following HEP analysis elements in its scope:
\begin{itemize}
\setlength\itemsep{0.1em}
    \item {\bf Combinations} --- A flexible way to describe combinations of objects to form new ones should be available. Examples include the reconstruction of all possible top quark candidates from the Lorentz boosted or resolved decay modes in an event; or a general way to express the combination of collections of objects to form a single collection of object tuples upon which algorithms can be applied. The main challenge is to guarantee type safety, i.e., to block users from combining uncombinable objects, which would be verified statically by the compiler. 
    \item {\bf Associations} --- The syntax to define one-to-many object associations, e.g., between a jet and its constituent particles or a track and its associated hits is required. The challenge is to devise a syntax that is clear and intuitive for the users and leaving the low-level routines to the compiler. 
    \item {\bf Low-level and non-standard objects} --- Support for defining low-level objects (e.g., hits, cells), and non-standard objects like long-lived particles (e.g., disappearing tracks, displaced muons, etc.) is needed. 
    \item {\bf Miscellaneous constructs} --- Support for handling multi-dimensional arrays and extended support for implicit loops  should be provided. This is needed for example, to loop through particles associated with a jet, or in cases where a selection is applied on a jet based on a quantity computed in comparison with every other jet in the same collection, such as ${\rm \min(\Delta R(jet, jet))}$. 
    \item {\bf Systematic uncertainties} --- A syntax is required to specify systematic uncertainties and describe how these are to be propagated to conclusions.      
\end{itemize}

\subsection{Advancing the infrastructures and auxiliary tools}

CutLang (and partially adl2tnm) already provide compiler infrastructures capable of handling the current ADL syntax and the foreseen near future additions to it, along with frameworks to perform basic analysis operations. However these tools have been built by physicists with no formal computing backgrounds, and could benefit from partial redesign based on formal computational methods.  Below we list several ideas that could advance the development of compiler  infrastructures for parsing and processing ADL and auxiliary tools that would contribute to extended uses of analyses written with a DSL syntax.  Some of these ideas are already being pursued for CutLang and adl2tnm.  They can also be adapted by other DSLs.

\begin{itemize}
\setlength\itemsep{0.1em}
    \item {\bf New compiler infrastructure:} --- A layered design can be implemented by recycling the existing functionality of CutLang and adl2tnm to a new framework and by partially reusing their frontend (e.g.,  parsing methods, \texttt{lex} and \texttt{yacc} modules, and translation rules). Performing semantic and logic checks on the ADL file is also needed. The specific goal would be to follow the LLVM~\cite{llvm} compiler infrastructure that provides its own LLVM Intermediate Representation (IR) language.
    Since some GPLs have already established a connection to LLVM, ADL would benefit from an integration with libraries written in these languages, and thus can be easily converted to those languages.

    \item {\bf Automated syntax verification:} --- This is required for a reliable use of the language. If one continues building on top of LLVM IR, a number of general purpose optimization passes and program verification tools~\cite{seahorn,DBLP:conf/vstte/MerzFS12,DBLP:conf/issta/WangZT15} which rely on state-of-the-art theorem provers and decision procedures such as~\cite{DBLP:conf/tacas/MouraB08} become available.
    Accepted practice in the verification community is to conduct verification  \emph{lazily} and make use of \emph{abstraction}~\cite{DBLP:conf/cav/ClarkeGJLV00,McMillan06}, but the choice of an appropriate abstraction is non-trivial and depends on the class of programs being verified.

    \item {\bf Interfacing with physics data types and tools:} --- HEP data is typically stored in ROOT files containing event-level information such as physics objects and their properties, and generic information, such as triggers, scale factors and weights. Any compiler for a DSL needs to be made mutually compatible and capable of automatically reading and processing a wide range of common data formats used by the LHC experiments and phenomenologists, in order to appeal to a broad user base. CutLang and adl2tnm have achieved this goal partially, but further work is required to reach complete automation.

    \item {\bf Static analysis:} --- The self-documenting nature of DSLs can be taken a step further with tools to assist and automate query among or comparison between multiple analyses in the space of event properties. Parsing source code to deduce facts about it without actually running the code is called \emph{static analysis}. 
    Preliminary work has been done to build tools to perform simple static analyses with ADL~\cite{Brooijmans:2020yij}. Such tools can help researchers get a complete view of which event final states are covered or not, which analyses have disjoint or overlapping selection regions, and inform researchers wishing to combine analyses or design new ones. 

    \item {\bf Differentiable programming:} --- A typical goal of an HEP analysis is to maximize quantities such as the expected statistical significance of a result. Such problems could be addressed by using differentiable programming provided that selection thresholds can be treated as differentiable parameters. A dedicated effort called {\tt GradHEP} has started within the HEP community towards building automatic differentiation tools to make analyses completely differentiable~\cite{gradhep}. The greatest challenge is to develop differentiable replacement analogues for non-differentiable operations such as binning and sorting commonly used in HEP analyses.  The DSL approach is well suited to serve as a medium for differentiable programming, as it systematically organizes the description of parameters to be differentiated.  Feasibility studies have started towards building an ADL-based differentiable analyses infrastructure for analysis optimization.

\end{itemize}

\subsection{Expanding practical use and physics applications}

ADL, or any other DSL for HEP analysis is only  meaningful if widely used in physics studies, which is feasible only if it fulfils a comprehensive set of analysis requirements. A critical step in achieving this is to put DSLs and their accompanying infrastructures to thorough practical use with a variety of physics implementations, even during the design phase. The ADL team has embraced this hands-on approach, which led to identifying and incorporating a number of missing features in the language and infrastructures. In order to further expand ADL's physics applicability, we intend to continue a direct involvement in physics studies through the activities listed below.  We invite the HEP community to participate in these activities and share feedback, which will help us provide an increasingly functional, robust and reliable language and infrastructures for HEP analysis, in particular, for realistic use with LHC data.

\begin{itemize}
\setlength\itemsep{0.1em}

    \item {\bf An analysis implementation and validation campaign:} --- Implementing a wide variety of published LHC analyses in ADL and keeping these synchronized with refinements to the language is essential and will be undertaken. Such implementations and debugging will likely continue to yield a list of unanticipated functionality needed to handle intricacies of real-world analyses. Moreover, community challenges, such as the HSF Data Analysis Working Group analysis benchmarks would also help to ensure the completeness of ADL or other DSLs. The syntactic consistency of implemented analyses could be verified using the automated methods described above. ADL algorithms could also be validated by comparing the results obtained from running ADL with those from the published analyses. To aid this process, we are building a validation chain for public use that automates the event production, analysis running and result comparison steps.

    \item {\bf A set of ADL hackathons:} --- Organizing a number of Hackathons on ADL analysis implementation and debugging will provide a focused and lively environment to effectively improve the functionality of ADL and CutLang.

    \item {\bf Studies for future colliders:} --- Physics sensitivity studies are crucial in motivating future colliders by demonstrating their physics capabilities. These studies, based on Monte Carlo events simulated with public tools are considerably less complex with respect to analysis of real data, and can already be performed with ADL and CutLang, as demonstrated in the published example~\cite{Paul:2020mul}.  We encourage the use of ADL and CutLang for such studies, which provide a practical analysis environment, even for colleagues with minimal analysis experience.

    \item {\bf Analysis of real data with ADL:} --- The ultimate goal of ADL is to accommodate all requirements of a detailed real data analysis in HEP.  ADL and CutLang are already partially employed in the ongoing design of two Run 2 ATLAS exotic searches, and more will follow both in ATLAS and CMS.  

\end{itemize}

\subsection{Analysis preservation}

The thousands of analyses designed by HEP experiments and phenomenologists covering a wide range of subjects offer a tremendous source of physics content that would serve to inspire new analysis ideas or train the next generation of physicists.  It is therefore critical to preserve this content in a most accessible and sustainable manner. The standard and self-documenting character of ADL renders it a naturally effective candidate for analysis preservation.  Below we list several ingredients for achieving analysis preservation with ADL.

\begin{itemize}

    \item {\bf  A set of ADL databases:} --- Three web-based, searchable, citable databases could be built to host content created with ADL in order to preserve analysis information. 
    First, an ADL analysis database would host ADL files of implemented analyses.  
    Second, an ADL objects database would host object definitions written in ADL (e.g., an ATLAS isolated tight electron or a CMS top quark identified with DNN-based medium tagging). Third, an ADL functions database would host external functions of non-trivial or non-analytical variables (e.g., complex kinematic variables, DNN discriminants).   Analysts would have access to object definitions or external functions by via unique identifiers, e.g., a {\tt doi} number, and editors could be built that allow navigating these unique identifiers. As the databases would be searchable, it would be possible to search for all analyses with 2 leptons or with missing transverse energy of at least 500 GeV, or for all non-isolated ATLAS muon definitions.  The HEP community would be able to add new analyses, functions, and objects to these databases. 

    \item {\bf LHC Run 2 and Run 3 analyses in ADL:} --- The new analyses using LHC Run 2 and Run 3 data could consider accompanying their publications with ADL implementations of their analysis algorithms as supplementary material.

    \item {\bf Documentation:} --- 
    Thorough documentation is indispensable for the wide-range use and sustainability of any construct or infrastructure, in particular those claiming to serve long-term analysis preservation.  Therefore, all information related to ADL including its formal definition will be clearly documented.  A user manual, tutorials, tools, analyses and physics studies will be easily accessible.   Source codes of the compiler infrastructures will be publicly available on so that interested parties can contribute to their  development.

\end{itemize}

\subsection{An AI vision for the future}

While our focus in this section has been to present prospects for the immediate future, our vision is a world in which the beginner student and the ageing professor use the same intuitive language to explore HEP data even long after the end of the experiment that produced them. It is also conceivable that the increasing power of artificial intelligence (AI) will establish direct human-machine interfaces that would respond to the spoken words of a future analyst:  ``get the ATLAS Run 2 data and simulation, select events with at least one lepton and 3 jets and missing transverse energy more than 100 GeV, display the transverse mass distribution, subtract contribution from top quark backgrounds, plot data versus simulation...", etc. Performing Big Data analyses this way would allow future generations of scientists to be more creative and industrious. 
Instead of spending time struggling with the syntax of a next-generation GPL, the future citizen scientist would spend time in an absorbing conversation with the AI interface to the LHC data. The ADL idea is a first step in realizing this vision. 

\section{Summary}

We presented the recently emerging approach of using a domain specific language (DSL) to express the physics algorithm of HEP data analysis in a standard and unambiguous way, via the specific example of Analysis Description Language (ADL). This approach increases the accessibility of the analysis physics algorithm by decoupling it from the analysis software frameworks, limiting the latter's scope only to technical operations. ADL, which is a declarative external DSL, can be rendered executable by any compiler infrastructure capable of handling its syntax, notably the runtime interpreter CutLang.  ADL can facilitate the abstraction, design, visualization, validation, combination, reproduction, (re)interpretation and overall communication of the physics algorithm, and  provide an organized and sustainable medium for long term analysis preservation.  Prototype design of ADL and CutLang along with several physics applications established the feasibility and potential of this approach.  A clear roadmap is laid out with concrete steps for future language extensions, technical advancements, physics studies and analysis preservation.  Pursuing it with the HEP community's support will evolve ADL into a robust, reliable and effective analysis environment that inspires innovative exploration of new ideas with HEP data towards scientific discovery.

\section*{Acknowledgements}

We thank our collaborators in the ADL/CutLang team for all their valuable contributions via technical or physics studies, and Grigory Fedyukovich for suggestions on the compiler infrastructures.  We also acknowledge the many ideas and support we received from colleagues who participated to the initial LHADA proposal.  
SS is supported by the Basic Science Research Program through the National Research Foundation of Korea (NRF) funded by the Ministry of Education under contracts NRF-2021R1I1A3048138, NRF-2018R1A6A1A06024970 and NRF-2008-00460.

\bibliographystyle{JHEP}
\bibliography{SnowmassADL}

\providecommand{\href}[2]{#2}\begingroup\raggedright\begin{thebibliography}{10}

\bibitem{Sekmen:2020vph}
S.~Sekmen, P.~Gras, L.~Gray, B.~Krikler, J.~Pivarski, H.B.~Prosper et~al.,
  \emph{{Analysis Description Languages for the LHC}}, {\emph{PoS} {\bfseries
  LHCP2020} (2020) 065} [\href{https://arxiv.org/abs/2011.01950}{{\ttfamily
  2011.01950}}].

\bibitem{BRUN199781}
R.~Brun and F.~Rademakers, \emph{{ROOT} — {A}n object oriented data analysis
  framework},
  \href{https://doi.org/https://doi.org/10.1016/S0168-9002(97)00048-X}{\emph{Nuclear
  Instruments and Methods in Physics Research Section A: Accelerators,
  Spectrometers, Detectors and Associated Equipment} {\bfseries 389} (1997) 81
  }.

\bibitem{FAST}
B.~Krikler, ``{FAST}.'' \url{https://fast-carpenter.readthedocs.io/en/latest/}.

\bibitem{NAIL}
A.~Rizzi, ``{NAIL}.''
  \url{https://indico.cern.ch/event/769263/timetable/#25-nail-a-prototype-analysis-l}.

\bibitem{Piparo:2699587}
D.~Piparo, P.~Canal, E.~Guiraud, X.~Valls~Pla, G.~Ganis, G.~Amadio et~al.,
  \emph{{RDataFrame: Easy parallel ROOT analysis at 100 threads}},
  \href{https://doi.org/10.1051/epjconf/201921406029}{\emph{EPJ Web Conf.}
  {\bfseries 214} (2019) 06029. 8 p}.

\bibitem{CMS:2020xwi}
{\scshape CMS Collaboration} collaboration, \emph{{Evidence for Higgs boson
  decay to a pair of muons}},
  \href{https://doi.org/10.1007/JHEP01(2021)148}{\emph{JHEP} {\bfseries 01}
  (2021) 148} [\href{https://arxiv.org/abs/2009.04363}{{\ttfamily
  2009.04363}}].

\bibitem{David:2021ohq}
P.~David, \emph{{Readable and efficient HEP data analysis with bamboo}},
  \href{https://doi.org/10.1051/epjconf/202125103052}{\emph{EPJ Web Conf.}
  {\bfseries 251} (2021) 03052}
  [\href{https://arxiv.org/abs/2103.01889}{{\ttfamily 2103.01889}}].

\bibitem{Proffitt:2021wfh}
M.~Proffitt and G.~Watts, \emph{{FuncADL: Functional Analysis Description
  Language}}, \href{https://doi.org/10.1051/epjconf/202125103068}{\emph{EPJ Web
  Conf.} {\bfseries 251} (2021) 03068}
  [\href{https://arxiv.org/abs/2103.02432}{{\ttfamily 2103.02432}}].

\bibitem{linq}
``{Language Integrated Query (LINQ)}.''
  \url{https://docs.microsoft.com/en-us/dotnet/csharp/programming-guide/concepts/linq/}.

\bibitem{Watts:2015saa}
G.~Watts, \emph{{Using Functional Languages and Declarative Programming to
  analyze ROOT data: LINQtoROOT}},
  \href{https://doi.org/10.1088/1742-6596/608/1/012024}{\emph{J. Phys. Conf.
  Ser.} {\bfseries 608} (2015) 012024}.

\bibitem{adlweb}
{H. B. Prosper, S. Sekmen and G. Unel}, ``{ADL} {W}eb {P}ortal.''
  \url{cern.ch/adl}.

\bibitem{Unel:2021edl}
G.~Unel, S.~Sekmen, A.M.~Toon, B.~Gokturk, B.~Orgen, A.~Paul et~al.,
  \emph{{CutLang V2: towards a unified Analysis Description Language}},
  \href{https://doi.org/10.3389/fdata.2021.659986}{\emph{Front. Big Data}
  {\bfseries 4:659986} (2021) }
  [\href{https://arxiv.org/abs/2101.09031}{{\ttfamily 2101.09031}}].

\bibitem{Prosper:2021umk}
H.B.~Prosper, S.~Sekmen, G.~Unel and A.~Paul, \emph{{Recent advances in ADL,
  CutLang and adl2tnm}},
  \href{https://doi.org/10.1051/epjconf/202125103062}{\emph{EPJ Web Conf.}
  {\bfseries 251} (2021) 03062}
  [\href{https://arxiv.org/abs/2108.00857}{{\ttfamily 2108.00857}}].

\bibitem{PartiQL}
J.~Pivarski, ``{P}arti{QL}.'' \url{https://github.com/jpivarski/PartiQL}.

\bibitem{awkward}
L.~Gray, ``{A}wkward{QL}.'' \url{https://github.com/lgray/AwkwardQL}.

\bibitem{Abdallah:2020pec}
{\scshape LHC Reinterpretation Forum} collaboration, \emph{{Reinterpretation of
  LHC Results for New Physics: Status and Recommendations after Run 2}},
  \href{https://doi.org/10.21468/SciPostPhys.9.2.022}{\emph{SciPost Phys.}
  {\bfseries 9} (2020) 022} [\href{https://arxiv.org/abs/2003.07868}{{\ttfamily
  2003.07868}}].

\bibitem{Kraml:2012sg}
S.~Kraml, B.C.~Allanach, M.~Mangano, H.B.~Prosper, S.~Sekmen, C.~Balazs et~al.,
  \emph{{Searches for New Physics: Les Houches Recommendations for the
  Presentation of LHC Results}},
  \href{https://doi.org/10.1140/epjc/s10052-012-1976-3}{\emph{Eur. Phys. J. C}
  {\bfseries 72} (2012) 1976}
  [\href{https://arxiv.org/abs/1203.2489}{{\ttfamily 1203.2489}}].

\bibitem{snowmass:preservation}
M.~Feickert, S.~Kranl et~al., \emph{{Data and Analysis Preservation, Recasting,
  and Reinterpretation}},  in \emph{Proceedings of the US Community Study on
  the Future of Particle Physics (Snowmass 2021)}, 2022.

\bibitem{Brooijmans:2016vro}
G.~Brooijmans, C.~Delaunay, A.~Delgado, C.~Englert, A.~Falkowski, B.~Fuks
  et~al., \emph{{Les Houches 2015: Physics at TeV colliders - new physics
  working group report}},  in \emph{{9th Les Houches Workshop on Physics at TeV
  Colliders (PhysTeV 2015) Les Houches, France, June 1-19, 2015}}, 2016
  [\href{https://arxiv.org/abs/1605.02684}{{\ttfamily 1605.02684}}].

\bibitem{Sekmen:2018ehb}
S.~Sekmen and G.~Ünel, \emph{{CutLang: A Particle Physics Analysis Description
  Language and Runtime Interpreter}},
  \href{https://doi.org/10.1016/j.cpc.2018.06.023}{\emph{Comput. Phys. Commun.}
  {\bfseries 233} (2018) 215}
  [\href{https://arxiv.org/abs/1801.05727}{{\ttfamily 1801.05727}}].

\bibitem{Unel:2019reo}
G.~Unel, S.~Sekmen and A.M.~Toon, \emph{{CutLang: a cut-based HEP analysis
  description language and runtime interpreter}},  in \emph{{19th International
  Workshop on Advanced Computing and Analysis Techniques in Physics Research:
  Empowering the revolution: Bringing Machine Learning to High Performance
  Computing (ACAT 2019) Saas-Fee, Switzerland, March 11-15, 2019}}, 2019
  [\href{https://arxiv.org/abs/1909.10621}{{\ttfamily 1909.10621}}].

\bibitem{Brooijmans:2018xbu}
G.~Brooijmans, M.~Dolan, S.~Gori, F.~Maltoni, M.~McCullough, P.~Musella et~al.,
  \emph{{Les Houches 2017: Physics at TeV Colliders New Physics Working Group
  Report}},  in \emph{{10th Les Houches Workshop on Physics at TeV Colliders
  (PhysTeV 2017) Les Houches, France, June 5-23, 2017}}, 2018,
  \href{http://lss.fnal.gov/archive/2017/conf/fermilab-conf-17-664-ppd.pdf}{http://lss.fnal.gov/archive/2017/conf/fermilab-conf-17-664-ppd.pdf}
  [\href{https://arxiv.org/abs/1803.10379}{{\ttfamily 1803.10379}}].

\bibitem{adllhcanl}
``{ADL LHC analyses repository}.''
  \url{https://github.com/ADL4HEP/ADLLHCanalyses}.

\bibitem{Paul:2020mul}
A.~Paul, S.~Sekmen and G.~Unel, \emph{{Down type iso-singlet quarks at the
  HL-LHC and FCC-hh}},
  \href{https://doi.org/10.1140/epjc/s10052-021-08982-4}{\emph{Eur. Phys. J. C}
  {\bfseries 81} (2021) 214}
  [\href{https://arxiv.org/abs/2006.10149}{{\ttfamily 2006.10149}}].

\bibitem{Adiguzel:2020brl}
A.~Adiguzel, O.~Cakir, U.~Kaya, V.E.~Ozcan, S.~Ozturk, S.~Sekmen et~al.,
  \emph{{Evaluating Analysis Description Language Concept as a First
  Introduction to Analysis in Particle Physics}},
  \href{https://arxiv.org/abs/2008.12034}{{\ttfamily 2008.12034}}.

\bibitem{llvm}
``{The LLVM Compiler Infrastructure}.'' \url{https://llvm.org/}.

\bibitem{seahorn}
A.~Gurfinkel, T.~Kahsai, A.~Komuravelli and J.A.~Navas, \emph{{The SeaHorn
  Verification Framework}},  in \emph{CAV}, vol.~9206, pp.~343--361, 2015.

\bibitem{DBLP:conf/vstte/MerzFS12}
F.~Merz, S.~Falke and C.~Sinz, \emph{{LLBMC: Bounded Model Checking of C and
  C++ Programs Using a Compiler IR}},  in \emph{VSTTE}, vol.~7152,
  pp.~146--161, 2012.

\bibitem{DBLP:conf/issta/WangZT15}
X.~Wang, L.~Zhang and P.~Tanofsky, \emph{Experience report: how is dynamic
  symbolic execution different from manual testing? a study on {KLEE}},  in
  \emph{Proceedings of the 2015 International Symposium on Software Testing and
  Analysis, {ISSTA} 2015, Baltimore, MD, USA, July 12-17, 2015}, M.~Young and
  T.~Xie, eds., pp.~199--210, {ACM}, 2015.

\bibitem{DBLP:conf/tacas/MouraB08}
L.M.~de~Moura and N.~Bj{\o}rner, \emph{{Z3: An Efficient SMT Solver}},  in
  \emph{TACAS}, vol.~4963, pp.~337--340, 2008.

\bibitem{DBLP:conf/cav/ClarkeGJLV00}
E.M.~Clarke, O.~Grumberg, S.~Jha, Y.~Lu and H.~Veith,
  \emph{Counterexample-guided abstraction refinement},  in \emph{CAV},
  vol.~1855, pp.~154--169, 2000.

\bibitem{McMillan06}
K.L.~McMillan, \emph{Lazy abstraction with interpolants},  in \emph{CAV},
  vol.~4144, pp.~123--136, 2006.

\bibitem{Brooijmans:2020yij}
G.~Brooijmans, A.~Buckley, S.~Caron, A.~Falkowski, B.~Fuks, A.~Gilbert et~al.,
  \emph{{Les Houches 2019 Physics at TeV Colliders: New Physics Working Group
  Report}},  in \emph{{11th Les Houches Workshop on Physics at TeV Colliders
  (PhysTeV 2019) Les Houches, France, June 10-28, 2019}}, 2, 2020
  [\href{https://arxiv.org/abs/2002.12220}{{\ttfamily 2002.12220}}].

\bibitem{gradhep}
``{gradHEP}.'' \url{https://gradhep.github.io/}.

\end{thebibliography}\endgroup

\end{document}